# Change of impervious space in German cities: A scaling law analysis

Rolf Bergs[1]

[1]PRAC Bergs & Issa Partnership Co.

___________________

**Research interest**

The leading research interest is to investigate trends of ground sealing in German cities organized as an own administrative district (kreisfreie Stadt) compared to those cities being part of a rural district (kreisabhängige Stadt). It is aimed to address a problem of the urban political economy of environment. We first explored the state of affairs in that specific research domain with a question to *chatgpt*: „Are there studies aimed to compare ground imperviousness among different types of German cities by using scaling law?" The AI generated answer was:

> „There have been several studies that have investigated the relationship between impervious ground cover and urban scaling laws in different types of cities in Germany. However, I'm not sure if there are any specific comparisons between different types of German cities with respect to impervious ground cover and urban scaling laws. …"

Since our research interest is exactly the latter aspect of the *chatgpt* reply, a particular exploration on that issue seemed reasonable.

**Hypothesis**

Due to substantially higher levels of gross value added, the cities with own administrative district (Administrative City Districts - ACD) are supposed to dispose of larger financial power for public investment as compared with a rural district. For the period 2014-2016, Hesse et al. (2019, p. 705) report tax revenues per capita in ADC of around 1,200 Euro on average while just 870 Euro for rural districts. Hence, ACDs dispose of around 40 percent higher tax revenues than rural districts. Even though cities within those districts (District Affiliated Cities - DAC) may have higher per capita tax revenues than smaller villages around, it is likely that there is still a substantial revenue gap as compared with ACDs. This may have a certain relevance for the urban environment and the climate problem since there is an obvious assumption that cities with more financial power may also invest larger volumes of public funds into urban infrastructure, eventually leading to more impervious space.

In terms of city population size, there is a larger overlap between both types of cities in the range between 35,000 and 3.6 million inhabitants (Fig. 1).

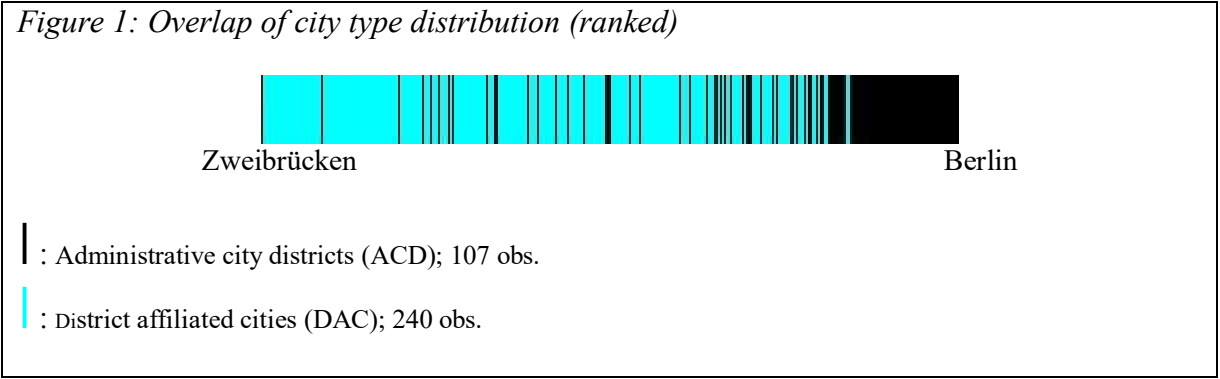

*Figure 1: Overlap of city type distribution (ranked)*

Zweibrücken             Berlin

| : Administrative city districts (ACD); 107 obs.

| : District affiliated cities (DAC); 240 obs.

Data source: Destatis (data file: Cities in Germany by area, population and population density 31 December 2021)

Hence, for Germany it is possible to directly compare the different levels and trends of urban ground sealing among different types of cities with similar size along the selection displayed by Fig. 1.

**Methodology**

Scaling law analysis (Bettencourt 2013) allows to (i) detect the proportion between population and sealed ground and (ii) the slope of the power distribution of sealed ground for the respective city distributions viewed. Methodologically, we distinguish between a sample of exclusively ACDs and another one with its lower tail replaced by DACs of comparable size. The time period of the study is 2006, 2009, 2012, 2015 and 2018. The information base of the study consists of ground imperviousness data provided by the IÖR-Monitor (www.ioer-monitor.de) in addition to official population and spatial city size data published by Destatis.

The idea for that research was inspired by a recent collection of papers addressing „Cities as Complex Systems" (Rybski and González 2022; Arcaute and Ramasco 2022)[1]. Urban scaling law is at the center of most those papers. It can be applied to show the changing proportion of any urban outcomes in relation to population size (e.g. production, sealed ground, pollution, political power or the rank in the urban system, like the famous Zipf's law), either for time series or cross-section analyses.

Urban scaling law is defined as:

$$Y \sim P^\beta$$

showing that any output $Y$ (a variable such as sealed ground) corresponds to the population size $P$ of a city, whereby the exponent $\beta$ indicates whether there is a convex or concave relationship for one city over time or within the distribution of all cities at one point of time.

The total impervious area of a sample of $n$ differently large cities with size $P_1$ to $P_n$ is thus:

$$Y = \sum_{i=1}^{n} \alpha P_i^\beta$$

with $\alpha$ and $\beta$ pre-estimated. Brelsford et al. (2020) (for the USA) and Bettencourt and Lobo (2015)[2] (for European countries) report β-estimates in the range between 0.81 (USA) and 1.09 (Spain). For German functional cities larger than 500,000 inhabitants authors found a β-estimate of 0.95. Globally, the β-exponent has shown to be around 5/6 for urban infrastructure as the above mentioned studies refer to.

The empirical analysis in our study comprises regression analyses for (i) the entire distribution of ACDs and (ii) a mix of upper tail ACDs merged with a lower tail DACs to explore a possible different pattern of ground sealing in both samples. In these regressions the simple comparison of those five years could disregard a possible temporal influence of earlier on later years. Therefore (iii), as proposed by Brelsford et al. (2020) we additionally use a *between* estimator:

$$\bar{y}_i = \alpha + \bar{p}_i \beta + u_i + \bar{\varepsilon}_i$$

where $u_i$ is the time-invariant individual effect.[3]

For the panel analysis, it is to be noted that during time ranks of city sizes have changed for a small number of cities, partly implied by territorial reforms during the period viewed. Therefore, some minor error might be induced in the estimations. The baseline year is 2006.

---

[1] Around 60 papers on that specific context are published in the Collection „Cities as complex systems", edited by Diego Rybski and Marta González.

[2] Bettencourt and Lobo (2015) use data on the urbanized area instead of impervious space.

[3] Brelsford et al. (2020) specify the between estimator with dummy variables for the years. Our results obtained with this method are almost identical but not reported.

# Results

Empirical results are structured into the simple power regressions per year regarded (Fig. 2 and Fig. 3) and the panel regressions to compare the elasticity of ground imperviousness in the homogenous ACD and the combined ACD-DAC sample (Table 1).

*Figure 2: All administrative city districts (ACDs) (n=107)*

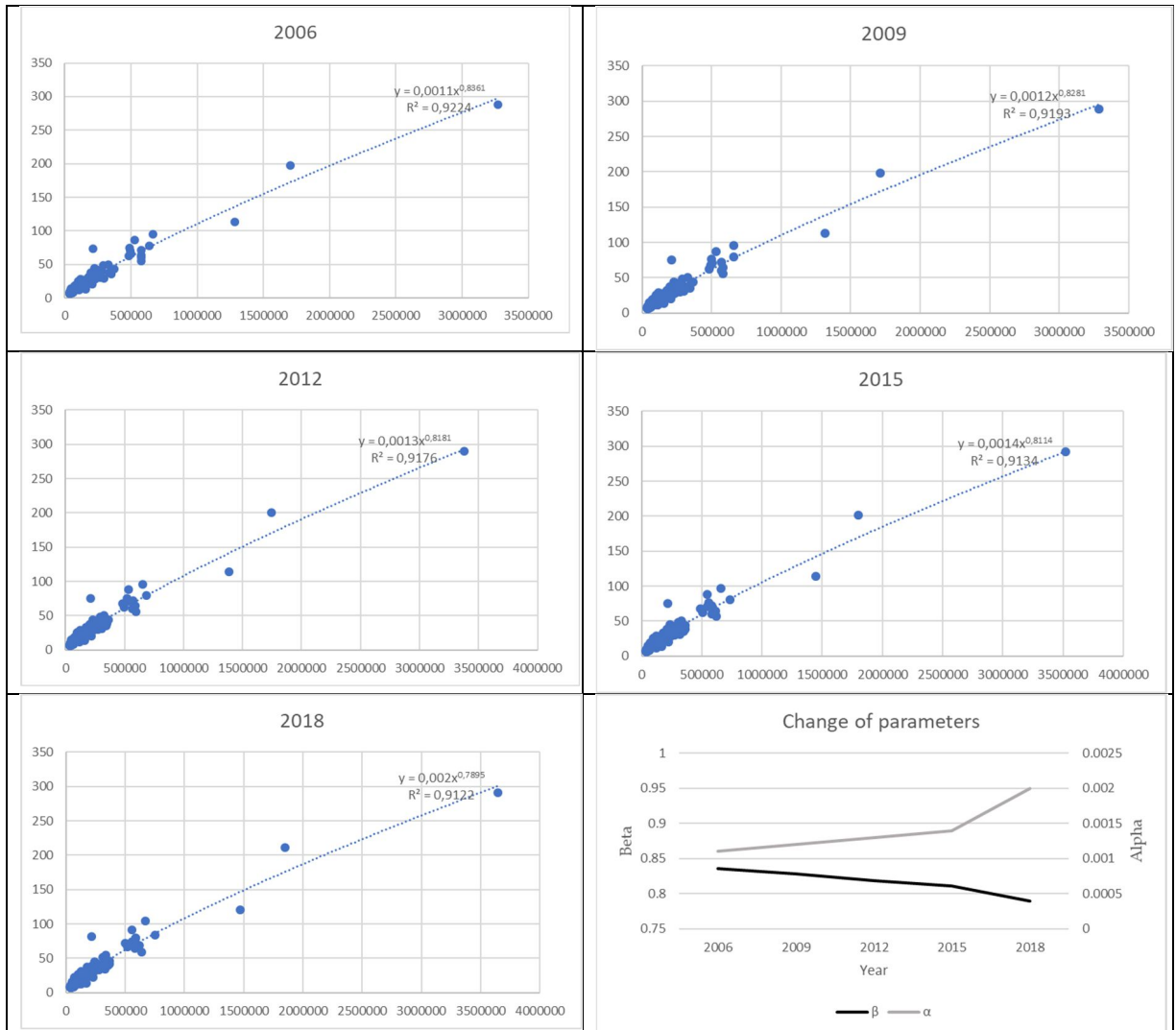

Unless otherwise stated: X-axis: city population *P*; Y-axis: sealed ground *Y* (square kilometers)
Data source: IÖR and Destatis

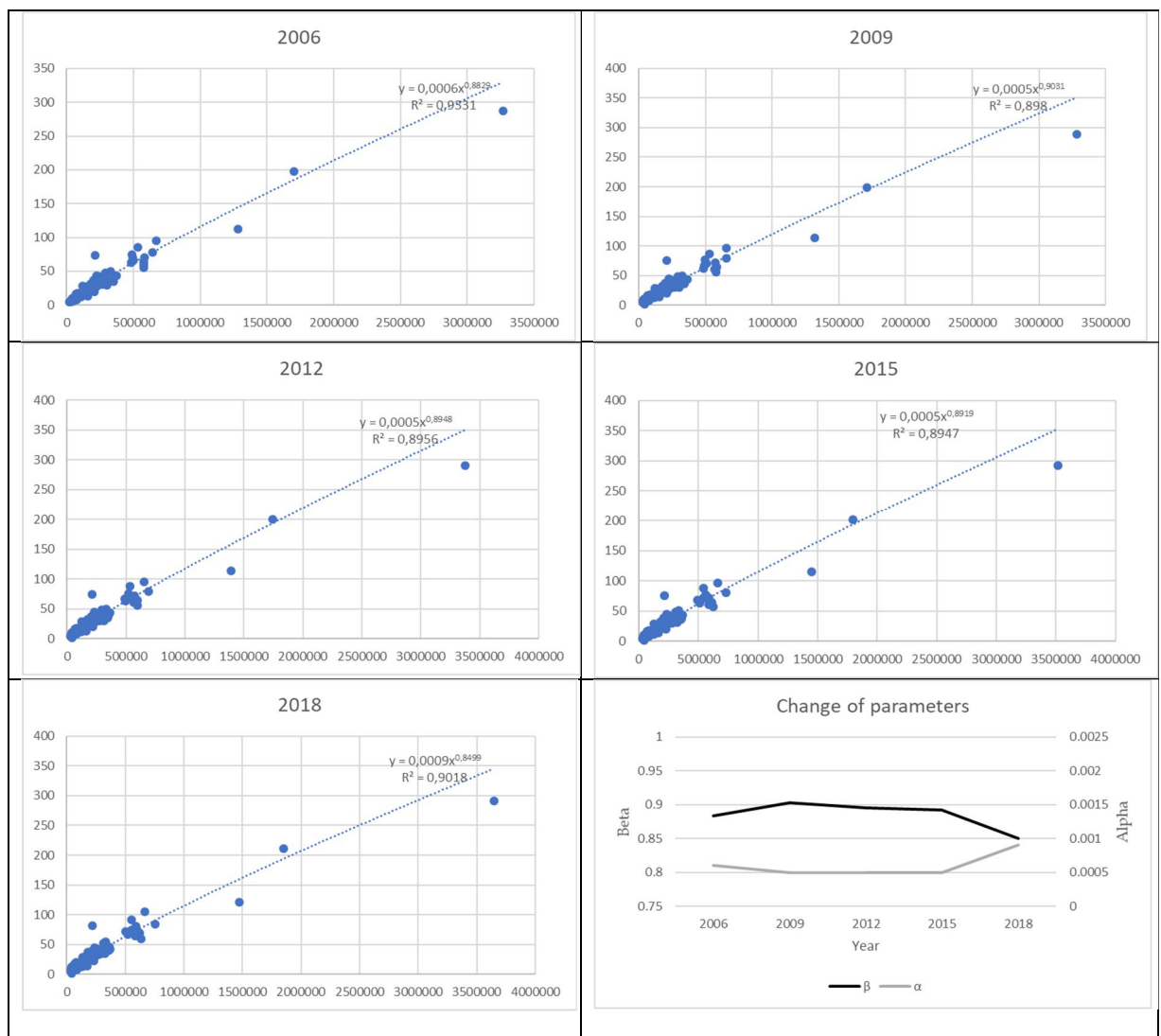

*Figure 3: Lower tail of ACDs replaced by district affiliated cities (DACs) of same size (n=106; n_DAC=47)*

Unless otherwise stated: X-axis: city population *P*; Y-axis: sealed ground *Y* (square kilometers)
Data source: IÖR and Destatis

*Table 1: Regression results for all administrative city districts (ACD) and for the sample with the replaced lower tail (ACD-DAC)*

| VARIABLES | (1) log (Y) | (2) log (Y) | (3) log (Y) | (4) log (Y) |
|---|---|---|---|---|
| log (P) | 0.817*** | 0.817*** | 0.885*** | 0.885*** |
|  | (0.0107) | (0.0189) | (0.0130) | (0.0218) |
| Constant | -6.598*** | -6.598*** | -7.463*** | -7.465*** |
|  | (0.128) | (0.225) | (0.159) | (0.259) |
|  |  |  |  |  |
| Observations | 535 | 535 | 530 | 530 |
| R-squared | 0.916 | 0.947 | 0.903 | 0.941 |
| Number of id |  | 107 |  | 106 |

Note: (1) Linear regression ACD; (2) Panel regression with between estimator (ACD); (3) Linear regression (ACD-DAC); (4) Panel regression with between estimator (ACD-DAC)
Standard errors in parentheses
*** p<0.01, ** p<0.05, * p<0.1

*Table 2: Sealed ground in ACDs 2018 and optimized scenario compared*

| Total sealed ground 2018 (square kilometers) | 3,585 |
|---|---|
| Sealed ground with revised parameters (square kilometers)* | 2,778 |
| Areal size difference (percent) | 22.52 |

Note: n=106 (smallest city was deleted)

* specified with: $Y = 0.0009 \cdot P^{\frac{5}{6}}$

Whether comparing different years and city samples or comparing different city samples in a panel approach, we find a significant difference between the sample of pure ACDs and one replaced by DACs for the lower tail. The non-linear (power) comparison of the two samples shows a more concave pattern for the sample with only ACDs while the sample with a replaced lower tail is closer to linearity. This reveals a relatively higher *per capita* imperviousness in the smaller ACDs as compared to the sample with replaced cities. The larger the ACDs are the lower appears to be their *per capita* imperviousness. Further to that, ground imperviousness has grown over time in the ACD sample while there was no such visible trend for the other sample. When looking at the panel analysis we recognize neglegible differences between the simple linear regression and the *between* estimates. This also corresponds with the results found by Brelsford et al. (2020) who explored US Metropolitan Statistical Areas based on census data. The elasticities estimated in that study are close to 5/6 and similar to our results. However, we could detect a stronger elasticity – i.e. estimates for *log(P)* - for the sample with ACDs replaced by DACs in the lower tail (i.e. an increase of one percent in population will add 0.89 percent in ground imperviousness, while it is just 0.82 percent for the sample with exclusively ACDs). This reveals both, a significantly lower level as well as a weaker trend of ground sealing over time for DACs alone. What is surprising at first glance, namely a higher elasticity for the hybrid ACD-DAC sample, is to be explained by the more skewed distribution of sealed ground. The base of the curve progression is significantly lower. The relevance of that can be demonstrated by comparing real ground imperviousness of the ACDs in 2018 with a function calibrated with the *a*-coefficient of the 2018 ACD-DAC regression and 5/6 as a global reference value for the exponent. This is exemplarily illustrated in Table 2. The result would be some impressive 22.52 percent less imperviousness in the ACD sample.

**Discussion**

The prior hypothesis of this study can thus be confirmed. However, significant differences in ground sealing levels and trends among different administrative types of cities with similar size, located in an economically integrated area like Germany, must have a reason, even if those differences are small. There are no visible natural reasons why agents in DACs seal less of their ground than their ACD counterparts do. Granting the status of an ACD or DAC is simply a political decision. If such a political decision has an implication on the level of local tax revenues it will automatically affect the expenditures from local budgets. Since there is a strong incentive of local administrations to attract and keep investment the size of the local budget is directly relevant for developing urban infrastructure. Zoning and land development to attract investment and labour essentially causes ground sealing in the city center and - more likely - along the urban fringes. Then, as put forward by Harvey and Clark some six decades ago, „Sprawl occurs, in fact, because it is economical in terms of the alternatives available to the occupants." (Harvey and Clark 1965). This simple rationale on the part of local authorities ignores the external environmental costs of ground sealing. In the end, the above results appear relevant for the urban political economy and the expenditure part of local public finance.[4]

---

[4] The results are specific for the German urban system. A relevant avenue of further research would be to investigate the relationship between local tax autonomy and urban ground imperviousness in national urban systems with other rules of local public finance.

———————

*17 April 2023*